\documentstyle[12pt]{article}
\begin{document}
\sf
\begin{flushright}
UCLA 93/TEP/31 \\ August 1993
\end{flushright}
\vspace{.5cm}
\renewcommand{\thefootnote}{\fnsymbol{footnote}}
\begin{center}
{\bf\sf MULTIPLE MUONS FROM NEUTRINO-INITIATED
MULTI-W(Z) PRODUCTION\footnote{Paper presented at 23rd Int.
Cosmic Ray Conf., July 21-30 1993, Calgary, Canada}
} \\
\quad   \\
D.A. Morris$^{1}$\footnote{email: morris@uclahep.physics.ucla.edu}
 and A. Ringwald$^{2}$\footnote{email: ringwald@cernvm.cern.ch}  \\
$^{1}$ University of California, Los Angeles, CA 90024, U.S.A. \\
$^{2}$ CERN, CH-1211, Geneva 23, Geneva, Switzerland
\\
  \quad
\end{center}

\begin{center}
ABSTRACT
\end{center}
Current underground detectors can search for
multiple muons from multi-W(Z) production initiated by
ultrahigh energy neutrinos from active galactic nuclei.
O($\mu$b) cross sections give rise to downward going
muon bundles whose features differ from those of
atmospheric muon bundles.

\vspace{-3\baselineskip}
\section*{}
1. INTRODUCTION
\vspace{\baselineskip}

\hspace*{\parindent}
A variety of recent theoretical results
has suggested the intriguing possibility that the cross section for
the nonperturbative production of O($\alpha_W^{-1}) \simeq 30$
weak gauge bosons (W,Z) may be as large as
$ O(100~\mbox{pb} - 10~\mu\mbox{b})$ above a parton-parton
center-of-mass threshold in the range 2.4--30~TeV
[Ri90,Es90,
Mc90,Co90,Ri91a].
Unfortunately, the theoretical evidence is largely
circumstantial and so it remains an open question as to
whether or not large cross sections for multi-W(Z) production are
realized in Nature. Though
the SSC, LHC and
Eloisatron can address this issue conclusively[Fa90,Ri91b],
it is natural to ask whether ultrahigh energy
cosmic rays can preemptively confront these conjectures.

\section*{}
\vspace{-3\baselineskip}
1.1  Proton and Neutrino Induced Interactions

\hspace*{\parindent}
If nonperturbative multi-W(Z) production exists, it can
be induced by energetic collisions between any two
weakly interacting partons ({\it e.g.,} $q,e,\nu$).
Characteristic byproducts of multi-W(Z) processes are
energetic prompt muons from W(Z) decays.
For example, if 30 W bosons are produced then
one can
expect $O( 30 \times \mbox{Br}(W\rightarrow \mu \nu_\mu) ) \simeq 3$
prompt muons which may be observed in deep underground detectors.

Multi-W(Z) production induced by cosmic protons is plagued by small
rates and poor signatures
due to competing generic processes with $O(40~\mbox{mb})$
cross sections[Mo93].
By contrast, multi-W(Z) production induced
by ultrahigh energy neutrinos competes only with
relatively small $O$(nb) charged-current reactions. If the
multi-W(Z) contribution to the neutrino-nucleon
total inelastic cross section $\sigma_{\mbox{tot}}^{\nu N}$
is also of $O$(nb), then near-horizontal muon bundles
provide a signature of neutrino-induced multi-W(Z) production
in the rock surrounding underground detectors[Mo91,Be92].
Large underwater detectors like DUMAND and NESTOR would also
be sensitive to such signals[Mo91,Be92,De92].
\nopagebreak[3] In this paper we broaden the prospects for detecting or
constraining neutrino-induced multi-W(Z) phenomena in underground
detectors by suggesting searches for neutrino-induced
muon bundles away from the horizontal direction.

\hspace*{\parindent}
To be quantitative we adopt a working hypothesis[Ri91b]
which parameterizes the sudden nonperturbative onset of
multi-W(Z) production in parton-parton subprocesses by
\begin{equation}
\hat \sigma_{\mbox{multi-W}} = \hat \sigma_0 \, \Theta
\bigl( \sqrt{\hat s} -\sqrt{\hat s_0 } \bigr).
\end{equation}
For purposes of illustration we will consider the
production of 30 W bosons by exploring parton-parton center-of-mass
thresholds in the range $\alpha_W^{-1} M_W \simeq 2.4~\mbox{TeV}
< \sqrt{\hat{s}_0} < 30~\mbox{TeV}$ and point cross sections
$ 100~\mbox{pb} < \hat{\sigma}_0 <  10~\mu\mbox{b}.$

\section*{}
\vspace{-3\baselineskip}
2. MULTI-W MUON BUNDLES INDUCED BY NEUTRINOS FROM AGN
\section*{}
\vspace{-3\baselineskip}
2.1  Constraints on Multi-W Phenomena

\hspace*{\parindent}
Apart from the speculative nature
of multi-W production, we must also contend with
a lack of knowledge of the flux of ultrahigh energy
cosmic neutrinos. A quark-neutrino center-of-mass
threshold of 2.4~TeV corresponds to
a neutrino energy of $\sim 3$~PeV where a
recent models have predicted
a sizeable neutrino flux from active galactic nuclei[St91,St92].
Regardless of any model, the Fly's Eye array puts upper limits[Ba85]
on the product of the flux
times total cross section for weakly interacting particles
(which we will assume are neutrinos)
in the range $10^8~\mbox{GeV} \leq E_\nu \leq 10^{11}~\mbox{GeV}$
if such particles initiate
extensive air showers deep in the atmosphere.
The limit applies only for
$\sigma_{\mbox{tot}}^{\nu N}(E_\nu) \le 10~\mu{\mbox{b}}$
since the possibility of flux
attenuation is neglected.

\pagestyle{plain}
\hspace*{\parindent}
Explicit parameterizations of the Fly's Eye limits, which we
denote by  $(j_\nu \sigma_{\mbox{tot}}^{\nu N})_{\mbox{FE}}(E_\nu),$
may be found in Refs. [Ma90,Mo91]. If one considers a particular
flux model $j_\nu^{\mbox{model}}(E_\nu)$
then in the $(E_\nu,\sigma_{\mbox{tot}}^{\nu N})$
plane the Fly's Eye excludes regions bounded by
\begin{equation}
10^8~{\mbox{GeV}}  <   E_\nu  <    10^{11}~{\mbox{GeV}}, \qquad
\displaystyle{
( j_\nu \sigma_{\mbox{tot}}^{\nu N})_{\mbox{FE}}(E_\nu)
               \over
j_\nu^{\mbox{model}}(E_\nu)}
 <  \sigma_{\mbox{tot}}^{\nu N}(E_\nu) < 10~\mu{\mbox{b}}.
\end{equation}
These inequalities may be translated into a
corresponding excluded region in
$(\sqrt{{\hat s}_0},\hat{\sigma}_0)$ space
which parameterizes multi-W phenomena. Fig.~1 shows
the excluded region of multi-W parameter space
for the (revised) flux of Stecker {\it et al.}[St91];
also indicated are the contours of
constant detection rates of multi-W muon bundles containing two
or more muons. These rates are integrated over all zenith angles
assuming standard muon energy-range relations with a detector depth
of 3700 hg/cm$^2$ and an idealized spherical Earth[Mo93].

\hspace*{\parindent}
For a $72~\mbox{m} \times 12~\mbox{m} \times 4.8~\mbox{m}$ detector
(MACRO) a vertical flux of 10$^{-13}~\mbox{ cm$^{-2}$ s$^{-1}$}$
corresponds to 26 events per year.
Consider two scenarios within reach
of such a detector: $\hat{\sigma}_0 = 10$~nb, 10~$\mu$b
for a common threshold of $\sqrt{\hat{s}_0} =$~2.4~TeV.
These cases correspond to total bundle detection rates of
$1.6\times10^{-15}$~\mbox{cm$^{-2}$ s$^{-1}$}
and
$3.2\times10^{-13}$~\mbox{cm$^{-2}$ s$^{-1}$} respectively.
As may be inferred from Fig. 2a, as $\hat\sigma_0$ increases,
the zenith angle distribution of muon bundles becomes less pronounced
in the near-horizontal direction and becomes more like the
distribution of background atmospheric bundles.
However, as seen in Fig. 2b, the relatively small pairwise separation
between multi-W muons may distinguish them from atmospheric
muons which have much larger separation[Be89,Ah92].

\begin{center}
\hspace*{0in}

\parbox{5in}{Fig. 1: Region of multi-W parameter space excluded (shaded) by
Fly's Eye assuming the flux of Stecker {\it et al.} [St91].
Dashed lines indicate constant multi-W muon bundle
flux (in cm$^{-2}$ s$^{-1}$) for detector depth of 3700~hg/cm$^{2}$.}
\end{center}

\hspace*{\parindent}Another feature of prompt muons from multi-W(Z)
processes is their large energy. The average muon energy
at the detector (depth 3700 hg/cm$^2$)
is 50 TeV (150 TeV) for $\hat\sigma_0 = 10~\mu \mbox{b}$
(10~\mbox{nb}).
Muons of this energy have a large probability
of undergoing catastrophic energy loss as they pass
through underground detectors[Al92, Me92].
In view of these characteristics, current underground
experiments should not constrain their searches for AGN
neutrinos to looking only in the near-horizontal direction:
they can also search for multi-W interactions by looking
for energetic, spatially compact muon bundles closer to
the zenith.

\hspace*{\parindent}
An additional technique for discriminating multi-W muon bundles from
generic muon bundles exploits the presence/absence of associated
extensive air showers. Surface arrays like those at EAS-TOP and
Soudan-II can furnish valuable information in this context.
Even for the largest $O(10~ \mu\mbox{b})$ cross sections we contemplate,
over 99\% of the corresponding vertical muon bundles
originate from multi-W interactions in the Earth. Hence energetic
muon bundles without an associated air shower provides an especially
convincing signature. The limiting factor in such searches
is the solid angle subtended by a surface array.

\begin{center}
\hspace*{0in}

\parbox{5.5in}{Fig. 2: Multi-W muon bundles detected at a depth of
3700 hg/cm$^2$ assuming the flux of Stecker {\it et al.} [St91].
Shown are curves for
$\hat{\sigma}_0 = 10$~nb (solid) and 10~$\mu$b (dashed)
for a common threshold of $\sqrt{\hat{s}_0} =$~2.4~TeV.
a) Zenith angle distribution of bundles
integrated with respect to $\cos\theta .$
b) Distribution of pairwise muon separation.
The solid histogram corresponds to normalized MACRO data (from two
supermodules) from Fig.~4 of Ref.~[Ah92].
Roughly 10\% of the bundles are dimuons and 90\% are trimuons.}
\end{center}

\section*{}
3. ACKNOWLEDGEMENTS
\vspace{\baselineskip}

\hspace*{\parindent}
We wish to acknowledge illuminating discussions with
M. Goodman and H. Meyer. D.A.M. is supported by the Eloisatron
project; he thanks the CERN theory group for its hospitality
while this work was being completed.

\section*{}
\vspace{-3\baselineskip}
REFERENCES
\vspace{\baselineskip}

\noindent Ahlen, S. {\it et al.} (MACRO): 1992, Phys. Rev., D46, 4836

\noindent Allison, W.W.M. {\it et al.} (Soudan II): 1992, Argonne preprint
ANL-HEP-CP-92-39

\noindent Baltrusaitis, R. {\it et al.} (Fly's Eye): 1985,  Phys. Rev., D31,
2192

\noindent Berger, Ch. {\it et al.} (Frejus): 1989, Phys. Rev., D40, 2163

\noindent Bergstr\"om L.,Liotta, R. and Rubinstein, H.: 1992,
 Phys. Lett., B276, 231

\noindent Cornwall, J.M.: 1990, Phys. Lett., B243, 271

\noindent Dell'Agnello L. {\it et al.}: 1992, INFN preprint DFF 178/12/1992

\noindent Espinosa, O.: 1990,  Nucl. Phys., B343, 310

\noindent Farrar, G.R. and Meng, R.: 1990, Phys. Rev. Lett., 65, 3377

\noindent MacGibbon, J.H. and Brandenberger, R.: 1990, Nucl. Phys., B331, 153

\noindent Meyer, H. (Frejus): 1992, Proc. of XXVIIth Recontre de Moriond,
Les Arcs, \\ \phantom{xxx} Gif-sur-Yvette, Ed. Frontieres, p. 169

\noindent McLerran, L.,Vainshtein, A. and Voloshin, M.:1990,
Phys. Rev., D42, 171

\noindent Morris, D.A. and Ringwald, A.: 1993, CERN preprint
CERN-TH.6822/93

\noindent Morris, D.A. and Rosenfeld, R.: 1991,  Phys. Rev.,
D44, 3530

\noindent Ringwald, A.: 1990, Nucl. Phys.,  B330, 1

\noindent Ringwald, A., Schrempp, F. and Wetterich, C.: 1991,
Nucl. Phys., B365, 3

\noindent Ringwald, A. and Wetterich, C.: 1991, Nucl. Phys.,
B353,  303

\noindent Stecker, F. {\it et al.}: 1991, Phys. Rev. Lett., 66,
2697; 1992: 69, 2738 (erratum)

\noindent Stenger, V.J.: 1992, DUMAND preprint DUMAND-9-92

\end{document}